\documentclass[preprintnumbers,amsmath,amssymb]{revtex4} 

\usepackage{graphicx}
\usepackage{dcolumn}
\usepackage{bm}

\usepackage{color}

\begin{document}

\title{Theoretical derivation of laser-dressed atomic states by using a fractal space}

\author{Guillaume Duchateau}
\affiliation{Universit\'e de Bordeaux-CNRS-CEA, Centre Lasers Intenses et Applications, UMR 5107, 351 Cours de la Lib\'eration, 33405 Talence Cedex, France}
\email{guillaume.duchateau@u-bordeaux.fr}

\date{\today}

\begin{abstract}
The derivation of approximate wave functions for an electron submitted to both a coulomb and a time-dependent laser electric fields, the so-called Coulomb-Volkov (CV) state, is addressed. Despite its derivation for continuum states does not exhibit any particular problem within the framework of the standard theory of quantum mechanics (QM), difficulties arise when considering an initially bound atomic state. Indeed the natural way of translating the unperturbed momentum by the laser vector potential is no longer possible since a bound state does not exhibit a plane wave form including explicitely a momentum. The use of a fractal space permits to naturally define a momentum for a bound wave function. Within this framework, it is shown how the derivation of laser-dressed bound states can be performed. Based on a generalized eikonal approach, a new expression for the laser-dressed states is also derived, fully symmetric relative to the continuum or bound nature of the initial unperturbed wave function. It includes an additional crossed term in the Volkov phase which was not obtained within the standard theory of quantum mechanics. The derivations within this fractal framework have highlighted other possible ways to derive approximate laser-dressed states in QM. After comparing the various obtained wave functions, an application to the prediction of the ionization probability of hydrogen targets by attosecond XUV pulses within the sudden approximation is provided. This approach allows to make predictions in various regimes depending on the laser intensity, going from the non-resonant multiphoton absorption to tunneling and barrier-suppression ionization. 
\end{abstract}


\maketitle

\section{Introduction}
For dozens of years and up to now, the interaction of intense femtosecond laser pulses with matter is investigated to understand the various physical mechanisms at play. In case of atomic targets, major fundamental processes are excitation, ionization, or harmonic generation, depending on the laser pulse characteristics as its peak intensity, duration, and central wavelength \cite{Corkum}. For not too intense laser pulses, i.e. the non relativistic case, the electron dynamics is theoretically described by the time-dependent Schr\"odinger equation (TDSE) where the influence of both the coulomb nucleus, of charge $Z$, and the laser electric fields are included. Theoretical predictions can be obtained through three main ways: by numerically solving the TDSE, by using perturbative approaches \cite{CohenTannoudjib, LandauQM}, or by searching for an analytical solution of the TDSE \cite{Reiss1970, Reiss1980, Faisal, Fabrice2015, Jain, Kaminski, Milosevic, Smirnova2008}. In the latter case, the laser-dressed atomic states can be described by the so-called Coulomb-Volkov (CV) wave functions which, depending on some physical parameters, correctly account for the electron dynamics, including multiphoton absorption and tunneling processes \cite{Reiss1970, Jain, Kaminski, Milosevic, D2, D3, D4, Arbo}. 

In the case of a purely free state ($Z = 0$), an unperturbed electron plane wave of momentum $\vec k$ (in atomic units) transforms into the Volkov wave function due to the external time-dependent laser electric field \cite{Volkov}. This state is an exact solution of the TDSE for $Z=0$. Starting from the unperturbed state, a technique to derive the laser-dressed one consists in translating $\vec k$ by the laser vector potential $\vec A(t)$, i.e. $\vec k$ transforms into the so-called generalized momentum $\vec p (t) = \vec k + \vec A(t)$. When a nucleus is present ($Z\neq 0$), the same concept of momentum translation can be used to construct a laser-dressed wave function from the initial unperturbed state \cite{Reiss1970, Reiss1980, Presnyakov}. Starting from unperturbed \textit{continuum} atomic states of positive energy which include a plane wave of momentum $\vec k$, the construction of CV states is obtained naturally by using the so-called momentum-translation approximation (MTA) \cite{Reiss1970, Reiss1980, Reiss1981}. Continuum CV states may also be derived more formally by looking for an approximate solution of the TDSE \cite{Reiss1970, Jain, D2}. Such a derivation is based on an ansatz specifying that the approximate wave function includes a Volkov phase, i.e. is the product of a Volkov phase and a function to be determined. The previous approaches have succeeded because the initial unperturbed wave function exhibits explicitely an initial momentum to which a laser-imposed translation can be applied, still dealing with the concept of MTA. In the case of initial \textit{bound} atomic state with negative energy, the wave function no longer includes explicitely a momentum which makes the MTA less efficient despite some attempts \cite{Reiss1989, Reiss1970, D2}. This observation, together with the fact that to find approximate solutions of the TDSE remains a challenging problem of interest for various applications, suggest that tackling this problem with an approach different from usual quantum-mechanical techniques could be beneficial to continue on progressing in the understanding of the laser-atom interaction.

Depending on various conditions, the interaction of an intense laser pulse with atomic targets exhibits a fractal behavior \cite{Gaier, Hillermeier, Main, Handke, Pohl}. This suggests that the use of a theoretical framework based on fractals could be well suited for the description of the dynamics of an electron submitted to both the coulomb and the time-dependent laser electric fields. Going further, space itself could be considered as exhibiting a fractal geometry. The concept of fractal space early appears following the introduction of Feynman's paths to describe the quantum behavior \cite{Feynman}. Within this description, the particle paths are very irregular and analogous to the Brownian motion which the trajectory exhibits a fractal structure, i.e. it exhibits the same pattern whatever the spatial scale \cite{Mandelbrot}. Actually, the TDSE can be obtained from the statistical mechanics of random walks \cite{Nelson} or equivalently by considering the classical Newtonian displacement of a particle on a fractal space \cite{Ord, nottale1993towards}. More generally, in the quest of providing an interpretation to the fundamental physical theories, like quantum mechanics (QM) \cite{auffeves2014contexts, Luchko, Stickler}, or to go further by deriving a consistent theory of quantum gravity, original approaches provide space or space-time with additional properties (in addition to the standard translation invariance, isotropy, differentiation, etc) as a fractal structure in particular \cite{Kroger2000, nottale1993towards, Calcagni2010, Calcagni2012, Calcagni2016}. Such approaches permit to derive fundamental equations as the Schr\"odinger, Klein-Gordon, or Dirac equations from first principles.

The previous considerations show the importance and the efficient mathematical usefulness of fractals in physical theories from fundamental aspects, where space-time itself is assumed to exhibit a fractal structure, to physical systems which the evolution intrinsically exhibit fractal behaviors. Based on previous works on fractal geometry, the main goal of the present work is to introduce the concept of fractal space to construct wave functions of electrons submitted to both the coulomb and a time-dependent laser electric fields. Since the fractal geometry may naturally lead to the TDSE \cite{Nelson, Feynman, nottale1993towards}, i.e. it is a geometric analogue of QM \cite{Ord}, it is shown hereafter to be well adapted for the present purpose, in particular to derive naturally bound CV states following the concept of MTA. It is worth mentioning that we do not claim the physical relevance of the fractal space postulate, despite it currently could not be excluded \cite{Kroger2000, Calcagni2010, Calcagni2012}. 

For the present purpose, the framework developed by Nottale and co-workers, the scale relativity theory (SRT), is used \cite{nottale1993towards, Nottale2013, celerier2004quantum, Nottale2009}. In the SRT, the mathematical structure of physical laws is assumed to be invariant under scale transformations and the assumption of space-time differentiation of any physical quantity is abandoned due to space fluctuations. A direct consequence is that the time derivative of the spatial position, i.e. the velocity, is no longer single-valued and leads to breaking of time reversal symmetry. To restore the latter symmetry, a generalized velocity with two vectorial components has been introduced, expressing as a complex number. The SRT has been shown to be well suited to describe bound unperturbed atomic states \cite{Nottale2013, agop2008fractal, Hermann1997}. In particular, this approach allows one to define a momentum for the bound states. It then becomes again possible to use naturally the concept of MTA to construct a bound CV wave function, in particular to use more formally the same ansatz as for continuum states relying on the dressing Volkov phase.

The paper is organized as follows. First, the considered physical problem and a particular solution are presented in Section \ref{sec:Statement} as a technical introduction, stressing on the key approximations, on which the further developments are performed. The framework of the SRT is used in Section \ref{sec:SRT}. The main concepts and equations used in the present work are firstly briefly reminded. Secondly, based on a first ansatz for the form of the wave function, a bound laser-dressed state is derived. Third, a more general ansatz based on the eikonal approach, suitable for both the continuum and bound CV states, is proposed. This derivation leads to a new expression of laser-dressed states which was not obtained within the standard QM. It includes the quiver electron motion which corresponds to the electron displacement only induced by the time-dependent laser electric field. This wave function thus carries more information on the interaction history compared to simplest versions of the CV states. It also pertains to the acceleration gauge where the TDSE is written in the quiver-moving reference frame. These derivations within the fractal space framework have paved us the way for further developments in the standard theory of QM as presented in Section \ref{sec:backToQM}. First, using an ansatz for the laser-dressed wave function form along with a development within the SRT, the quantum mechanical derivation of a bound CV state is made. Due to the above-mentioned inherent problem of using the MTA for a bound wave function, the demonstration is however not analogous to the continuum case and includes a higher level of assumptions. Second, a more general CV state including an additional phase (compared to the simplest CV states) also pertaining to the acceleration gauge is derived.
In Section \ref{sec:discussion}, a comparison of the various laser-dressed states is done, stressing in particular on the link between the wave functions including the quiver motion and the sudden approximation in order to define a first simple application of these states. In order to evaluate whether the latter wave functions are able to make physically reasonable predictions and also check their reliability, the ionization of hydrogen targets by attosecond XUV pulses within the sudden approximation is addressed. These wave functions including the quiver motion allow to recover correct results while the simplest CV states predict no transition. The present approach allows to make predictions in various regimes depending on the laser intensity, going from the non-resonant multiphoton absorption to tunneling and barrier-suppression ionization. Finally, conclusions are drawn together with the perspectives of the present work in Section \ref{sec:conclusion}.

Atomic units ($e = m = \hbar = 1$) are used throughout the paper unless otherwise stated.

\section{Statement of the problem}
\label{sec:Statement}
The single electron dynamics in both the coulomb and the laser electric fields is described by the electron wave function $\Psi (\vec{r},t)$ which evolution is driven by the TDSE:
\begin{equation} 
i\frac{\partial \Psi (\vec{r},t)}{\partial t} = \left( -\frac{\nabla^{2}}{2} + V_c + V_L \right) \Psi (\vec{r},t)
\label{eq:TDSE}
\end{equation} 
where $V_c = -Z/r$ is the coulomb potential with $Z$ the charge of the nucleus, and $V_L = \vec{r}.\vec{E}(t)$ is the interaction Hamiltonian with the laser electric field $\vec{E}(t)$ in the length gauge and within the electric dipolar approximation. In case of $Z=0$, i.e. a free electron moving in an oscillating electric field, Eq. (\ref{eq:TDSE}) admits an exact solution, the Volkov state, which reads \cite{Volkov}:
\begin{equation} 
\vartheta_{\vec k} (\vec{r},t) = \frac{1}{(2 \pi)^{3/2}} \exp \left( i\vec{p}(t).\vec{r} - \frac{i}{2} \int_0^t dt' p^{2}(t') \right)
\label{eq:Volkov}
\end{equation} 
where $\vec{p}(t) = \vec{k} + \vec{A}(t)$ is the canonical momentum, with $\vec{k}$ the field-free momentum of the electron and $\vec{A}(t)$ the laser vector potential related to the laser electric field as $\vec{E}(t) = - \partial \vec{A}(t) / \partial t$. This expression of the canonical momentum is linked to the above-mentioned MTA, and is obtained by integrating $\partial \vec{p}(t) / \partial t = - \vec{E}(t)$ which accounts for the fundamental principle of classical dynamics where the electron dynamics is only driven by the laser electric field.

In case of $Z \neq 0$, there is no exact solution of Eq. (\ref{eq:TDSE}) \cite{Reiss1970, Reiss1980, Jain}. Two classes of solutions may be defined: laser-dressed continuum and bound wave functions, $\Psi_f^-(\vec{r},t)$ and $\Psi_i^+(\vec{r},t)$, respectively. These standard notations belongs to the scattering theory of collisions where lowerscripts $f$ and $i$ refer to final and initial wave function, respectively, and upperscripts $-$ and $+$ denote incoming and outgoing waves, respectively. Initial conditions are $\lim\limits_{t\rightarrow \infty} \Psi_f^-(\vec{r},t) = \Phi_{\vec{k}}^-(\vec{r},t)$ and $\lim\limits_{t\rightarrow -\infty} \Psi_i^+ (\vec{r},t) = \Phi_{i}(\vec{r},t)$ where $\Phi_{\vec{k}}^-(\vec{r},t)$ and $\Phi_{i}(\vec{r},t)$ are stationary unperturbed continuum (with positive eigenenergy) and bound (with negative eigenenergy) states, respectively, i.e. exact solutions of Eq. (\ref{eq:TDSE}) without external field ($V_L  = 0$). The notation $\Phi_0(\vec{r},t) = \varphi_0(\vec{r})\exp (-iE_0 t)$ is used throughout the paper, where $E_0$ denotes the energy of the unperturbed state. $E_0 = k^2 / 2$ and $E_0 < 0$ for continuum and bound states, respectively.

In case of laser-dressed continuum states, an approximate solution of Eq. (\ref{eq:TDSE}), $\chi_f^-(\vec{r},t)$, can be derived by assuming that the Volkov phase imposes the main temporal evolution of the wave function in conditions of intense laser field \cite{D2}:
\begin{equation} 
\chi_f^-(\vec{r},t) = f^-(\vec{r},t) \times \exp \left( i\vec{p}(t).\vec{r} - \frac{i}{2} \int_\infty^t dt' p^{2}(t') \right)
\label{eq:chif}
\end{equation} 
where $f^-(\vec{r},t)$ is a function to be determined, and $\vec{p}(t)$ is unknown \textit{a priori}. Introducing the expression (\ref{eq:chif}) in Eq. (\ref{eq:TDSE}) leads to:
\begin{equation} 
\left( i\frac{\partial}{\partial t} - \frac{\partial \vec{p}(t)}{\partial t}.\vec r \right) f^-(\vec{r},t) = \left( -\frac{\nabla^{2}}{2} - \frac{Z}{r} -i \vec{p}(t).\vec{\nabla} + \vec{r}.\vec{E}(t) \right) f^- (\vec{r},t)
\label{eq:dfdt}
\end{equation}
Assuming that the electron dynamics is mainly driven by the laser electric field, the expression of $\vec{p}(t)$ can be obtained from the fundamental principle of classical dynamics:
\begin{equation} 
\frac{\partial \vec{p}(t)}{\partial t} = - \vec{E}(t)
\label{eq:dpdt}
\end{equation}
which solution is $\vec{p}(t) = \vec{k} + \vec{A}(t)$ with $\vec{A}(t) = - \int_{\infty}^{t} dt' \vec E (t')$ and the initial condition $\vec{p}_0 = \vec{p}(\infty) = \vec{k}$. Note that in the case of initial bound states, the difficulty arises here from the fact that the initial momentum $\vec{p}_0$ is not defined. Further assuming that the temporal variations of $f^-(\vec{r},t)$ are mainly due to the external field, Eq. (\ref{eq:dfdt}) can be split into two equations:
\begin{eqnarray} 
\label{eq:dfdtsplitted}
i\frac{\partial f^-(\vec{r},t)}{\partial t} = - i \vec{A}(t).\vec{\nabla} f^-(\vec{r},t) \\
\left( -\frac{\nabla^{2}}{2} - \frac{Z}{r} -i \vec{k}.\vec{\nabla} \right) f^- (\vec{r},t) = 0
\label{eq:dfdtsplittedn}
\end{eqnarray}
If the interaction is short enough, the temporal variations of $f^-(\vec{r},t)$ may be neglected \cite{D2}, i.e. the crossed term $\vec{A}(t).\vec{\nabla} f^-(\vec{r},t)$ is neglected. The approximation is also correct for long pulses with not too large electric field amplitudes \cite{D2}.
It follows that $f^- (\vec{r},t)$ is nothing but a confluent hypergeometric function $_1F_1(iZ/k,1,ikr-i\vec k.\vec r)$ accounting for the coulomb influence of the nucleus on the electron dynamics. Since $\varphi_{\vec{k}}^- (\vec{r}) = f^-(\vec{r})\exp (i\vec{k}.\vec r)/(2\pi)^{3/2}$ and $E_0 = k^2 / 2$, the approximate solution of Eq. (\ref{eq:TDSE}) reads:
\begin{eqnarray} 
\chi_f^{-(1)}(\vec{r},t) & = \varphi_{\vec{k}}^- (\vec{r}) \times \exp \left( i\vec{A}(t).\vec{r} - \frac{i}{2} \int_\infty^t dt' p^{2}(t') \right) \\ 
& = \Phi_{\vec{k}}^- (\vec{r},t) \times \exp \left( i\vec{A}(t).\vec{r} - i\vec{k}.\vec{\alpha}(t)  - \frac{i}{2} \int_\infty^t dt' A^{2}(t') \right) \nonumber
\label{eq:solchi}
\end{eqnarray}
where $\vec{\alpha}(t) = \int_{\infty}^t dt' \vec{A}(t')$ corresponds to the electron quiver motion, i.e. the electron displacement only induced by the laser electric field. $\chi_f^-$ is called a CV state due to its form mixing an unperturbed coulomb state and a Volkov phase, and corresponds here to one of the simplest forms.

\section{Laser-dressed atomic states in a fractal space}
\label{sec:SRT}
\subsection{Theoretical framework}
The following general SRT framework has been introduced by Nottale and co-workers \cite{nottale1993towards, Nottale2013, celerier2004quantum}. Here is provided a sketch of the main concepts for the present purpose of constructing laser-dressed atomic states. Since the assumption of usual differentiation must be abandoned in a fractal space (the variations of a quantity depends on the scale), the differential operator must be redefined \cite{Mandelbrot}. An infinitesimal displacement reads $d\vec R = d\vec r + d\vec\xi$ where $d\vec r = \vec v dt$ corresponds to a continuously differentiable space, and $d\vec \xi = \vec a \sqrt{dt}$ accounts for the fractal structure. In the latter expression, $\vec a$ is a characteristic constant vector and the fractal dimension $D_H=2$ is choosen such that the standard quantum mechanical behavior is recovered as shown hereafter. Note that this value of the fractal dimension is responsible for the specific evolution of $d\xi$ as $\sqrt{dt}$. The expression of the velocity, which is the temporal variation of position, thus has to be redefined. For a given scale, due to the spatial fluctuations of space, the velocity is no longer continuous and exhibits two values around a given position. Indeed, defining $\vec v_- = (\vec r(t)-\vec r(t-dt))/dt = d\vec r / dt_-$ and $\vec v_+ = (\vec r(t+dt)-\vec r(t))/dt = d\vec r / dt_+$, $\vec v_- \neq \vec v_+$ in general. The temporal reversibility, where the change $dt \rightarrow -dt$ leaves fundamental equations invariant, is then broken. A two-value velocity at a given position can be naturally described by complex numbers from which a differential operator can be defined to restore the temporal symetry:
\begin{equation} 
\frac{\hat d}{dt} = \frac{1}{2}\left( \frac{d}{dt_+} + \frac{d}{dt_-} \right) - \frac{i}{2}\left( \frac{d}{dt_+} - \frac{d}{dt_-} \right)
\end{equation}
The generalized velocity then reads:
\begin{equation} 
{\vec{\cal V}} = \frac{\hat d {\vec{r}}}{dt} =  \frac{1}{2}\left( {\vec v_+} + {\vec v_-} \right) - \frac{i}{2}\left( {\vec v_+} - {\vec v_-} \right) = \vec V - i \vec U 
\end{equation}
For a non-fractal (continuous) space structure with $\vec v_- = \vec v_+ = \vec v$, such an expression provides the expected standard behavior: $\vec V=\vec v$ and $\vec U=\vec 0$. In the general case, an imaginary part of the velocity emerges from the fractal structure of space. Within such conditions ($D_H=2$ in particular), it can be also shown that the temporal differential operator reads \cite{nottale1993towards, Nottale2013}:
\begin{equation} 
\frac{\hat d}{dt} = \frac{\partial}{\partial t} + {\vec{\cal V}.\vec \nabla} - \frac{i}{2} \Delta
\label{eq:ddt}
\end{equation}
where the operators $\vec \nabla$ and $\Delta$ are the standard spatial gradient and laplacian operators, respectively. By generalizing the lagrangian formalism and least action principle to the present theoretical framework, Euler-Lagrange equations can be derived, from which a generalized fundamental equation for the system dynamics induced by an external potential $V_{ext}$ can be obtained:
\begin{equation} 
\frac{\hat d {\vec{\cal V}}}{dt} = - {\vec{\nabla}} V_{ext}
\label{eq:Newton}
\end{equation}
where we remind that the electron mass is set to unity in atomic units ($m=1$). For the present purpose, $V_{ext}$ should account for both coulomb and laser influences. The generalized momentum $\vec{\cal P}$ is linked to the generalized action $\cal S$ as ${\vec{{\cal P}}} = {\vec{\nabla}} \cal S$. By using Eq. (\ref{eq:ddt}), a spatial integration of Eq. (\ref{eq:Newton}) leads to:
\begin{equation}
\frac{\partial \cal S}{\partial t} - \frac{i}{2}\vec \nabla .{\vec{\cal V}} + \frac{1}{2} {\vec{\cal V}}^2 + V_{ext} = 0
\label{eq:Newton2}
\end{equation}
where the boundary condition is taken at infinity (all terms are zero since no particle is present and the potential vanishes). A new function $\Psi$ can also be introduced such that $\Psi = \exp (i \cal S )$ as introduced by Feynman \cite{Feynman}. Since ${\vec{{\cal P}}} = {\vec{{\cal V}}}$ ($m=1$), the generalized velocity may be expressed as a function of $\Psi$ as ${\vec{{\cal V}}} = {\vec{\nabla}} {\cal S} = -i {\vec{\nabla}} \ln \Psi$. The introduction of the latter expression in Eq. (\ref{eq:Newton2}) leads to the TDSE where $\Psi$ might be identified to the wave function of QM as defined by Schr\"odinger.

For the present purpose, let us now consider a stationary coulomb state without external field, with associated generalized velociy ${\vec{\cal V}}_0$ which does not depend on time. In that case, Eq. (\ref{eq:Newton2}) transforms into:
\begin{equation}
- \frac{i}{2}\vec \nabla .{\vec{\cal V}}_0 + \frac{1}{2} {\vec{\cal V}}_0^2 + V_{c} = 0
\label{eq:NewtonStationnary}
\end{equation}
Note that in case of a stationary bound state, ${\vec{\cal V}}_0$ corresponds to a purely imaginary velocity field \cite{Hermann1997, Nottale2013}. The zero real part of the velocity is consistent with an interpretation of QM as presented by Bohm \cite{Bohm}. Indeed, a bound state may be seen as a particle at rest in average. Its wave function may generally be written as:
\begin{equation}
\Phi_0 (\vec r , t) = \exp \left( i \int d\vec r . {\vec{\cal V}}_0 -i E_0 t\right)
\label{eq:unperturbedstate}
\end{equation}
where the spatial integration is such that $\vec \nabla \int d\vec r . {\vec{\cal V}}_0 = {\vec{\cal V}}_0$. For instance, in case of the hydrogen 1s state which the wave function is purely exponential, ${\vec{\cal V}}_0 = i \hat{r}$ where $\hat{r}$ is the unitary vector in the direction of $\vec r$ in spherical coordinates. In that case, the stationary bound state reads $\varphi_0 (\vec r) \propto \exp ( i {\vec{\cal V}}_0.\vec r)$.

\subsection{Derivation for laser-dressed bound states}
Based on the previous considerations, a laser-dressed bound state can be derived as follows. As for the previous derivation under the framework of the standard theory of QM, an ansatz on the form of the wave function is required. As a first attempt, along with the quantum mechanical ansatz (\ref{eq:chif}) and the SRT form of an unperturbed state (\ref{eq:unperturbedstate}), we choose:
\begin{equation}
\chi_i^+(\vec{r},t) = \exp \left( i \int d\vec r \ {\vec{\cal W}}(\vec{r},t) - i \int_{-\beta t}^t dt' {\cal E}(t') \right)
\label{eq:ansatzChii}
\end{equation}
where now ${\vec{\cal W}}$ depends on both space and time, and ${\cal E}(t)$ is the time-dependent electron energy; ${\vec{\cal W}}$ and ${\cal E}(t)$ having to be determined. As an initial condition, $\chi_i^+(\vec{r},t)$ should be equal to $\Phi_i (\vec{r},t) = \Phi_0 (\vec{r},t)$ before any interaction, i.e. for $t<0$. It follows that ${\vec{\cal W}}(\vec{r},t<0) = {\vec{\cal V}}_0$ and ${\cal E}(t \leq 0) = E_i$. Note that the lower bound on the temporal energy integration, $-\beta t$ with $\beta \geq 0$, has been introduced to satisfy the required initial condition. $\beta \rightarrow 0$ will be imposed by the end of the derivation. In order to determine ${\vec{\cal W}}(\vec{r},t)$ and ${\cal E}(t)$, we may write Eq. (\ref{eq:Newton2}) with ${\cal S} = \int d\vec r \ {\vec{\cal W}}(\vec{r},t) - \int_{-\beta t}^t dt' {\cal E}(t')$ coming from Eq. (\ref{eq:ansatzChii}), leading to:
\begin{equation}
-{\cal E}(t) + {\cal E}(-\beta t) + \int d\vec r . \frac{\partial {\vec{\cal W}}}{\partial t} - \frac{i}{2}\vec \nabla .{\vec{\cal W}} + \frac{1}{2} {\vec{\cal W}}^2 + V_{c} + \vec{r}.\vec{E}(t) = 0
\label{eq:TDSEchii}
\end{equation}
Assuming, as previously, that the temporal velocity variations are mainly due to the external electric field, they are given by:
\begin{equation}
\frac{\partial {\vec{\cal W}}}{\partial t} = - \vec{E}(t)
\end{equation}
which integration provides ${\vec{\cal W}} = {\vec{\cal V}}_0 + \vec{A}(t)$ with $\vec{A}(t) = - \int_{0}^{t} dt' \vec E (t')$. The MTA can be used here because the momentum (velocity) of the initial bound state can be defined properly within the SRT framework. Eq. (\ref{eq:TDSEchii}) then transforms into:
\begin{equation}
-{\cal E}(t) + {\cal E}(-\beta t) + \frac{A(t)^2}{2} + {\vec{\cal V}}_0.\vec{A}(t) - \frac{i}{2}\vec \nabla .{\vec{\cal V}}_0 + \frac{1}{2} {\vec{\cal V}}_0^2 + V_{c} = 0
\label{eq:TDSEchii_2}
\end{equation}
Using Eq. (\ref{eq:NewtonStationnary}), it leads to:
\begin{equation}
-{\cal E}(t) + {\cal E}(-\beta t) + \frac{A(t)^2}{2} + {\vec{\cal V}}_0.\vec{A}(t) = 0
\label{eq:TDSEchii_3}
\end{equation}
where the crossed term ${\vec{\cal V}}_0.\vec{A}(t)$ depends on both space and time. Since ${\cal E}(t)$ only depends on time and following the same level of approximation as in the previous section, ${\vec{\cal V}}_0.\vec{A}(t)$ should be removed from Eq. (\ref{eq:TDSEchii_3}) by consistency, leading to ${\cal E}(t) = {\cal E}(-\beta t) + A(t)^2/2$. By setting $\beta$ to zero (${\cal E}(0) = E_i$) to ensure the continuity in the wave function at the time when the laser pulse is switched on, the bound CV state then reads:
\begin{eqnarray}
\label{eq:Chii}
\chi_i^+(\vec{r},t) & = \exp \left( i \int d\vec r \ ( {\vec{\cal V}}_0 + \vec{A}(t) ) - i E_i t - \frac{i}{2} \int_0^t dt A(t)^2 \right) \\
& = \Phi_0(\vec r , t) \exp \left( i \vec{A}(t).\vec r - \frac{i}{2} \int_0^t dt A(t)^2 \right) \nonumber
\end{eqnarray}
where Eq. (\ref{eq:unperturbedstate}) has been used. This state is hereafter called $\chi_i^{+(1)SRT}(\vec{r},t)$.

The previous calculations have shown that a bound CV state can be derived assuming a fractal structure of space, with the same level of approximation as in the standard theory of QM, i.e. neglecting a crossed term as in Eq. (\ref{eq:dfdtsplitted}). That shows the reliability of the ansatz (\ref{eq:ansatzChii}) involving the energy of the initial state, suggesting it could be used in standard QM. This is done hereafter.

Note that such an approach based on the ansatz (\ref{eq:ansatzChii}) could be used for a continuum state. However, no crossed term in the Volkov phase of the CV wave function appears within that condition. This is due to the independence of both arguments of the exponential function in the expression (\ref{eq:ansatzChii}), where in particular the temporal energy part is fully general instead of being a function of the momentum appearing in the spatial part. This consideration underlines the importance of the ansatz choice and motivates the following.

\subsection{General derivation}
For the sake of comparison, the previous derivation has been performed by postulating an ansatz on the wave function form along with previous works in the framework of the standard theory of QM. Under the SRT framework, another ansatz, more natural and elegant whatever the ingoing or outgoing characteristic of the wave function, consists in simply expressing the laser-dressed state as a function of the generalized action, $\chi(\vec{r},t) = \exp (i {\cal S}(\vec{r},t) )$, i.e. the most general exponential form (instead of providing some expressions in the argument of the exponential as in Eq. (\ref{eq:ansatzChii})). Such an approach which pertains to the eikonal approximation, i.e. the WKB quasi-classical approximation, was also used in QM as in \cite{Gersten1975, Smirnova2006, Smirnova2008} for instance. Within the present framework, that takes a particular significance since the electron dynamics can be described by the classical Newton-like equation (\ref{eq:Newton}), or the equivalent Eq. (\ref{eq:Newton2}) which is now used to determine an expression of the generalized action ${\cal S}$.

Still assuming that the temporal variations of the velocity field ${\vec{{\cal V}}}$ are mainly due to the laser electric field, $\partial {\vec{\cal V}} / \partial t = - \vec{E}(t)$ is stated (which corresponds to a generalized MTA), i.e. ${\vec{\cal V}} = {\vec{\cal V}}_0^{\pm} + \vec{A}(t)$ with $\vec{A}(t) = - \int_{t_0}^{t} dt' \vec E (t')$. Depending on the nature of the initial state, bound or continuum, the initial condition is taken for $t_0 = -\infty$ or $t_0 = \infty$, respectively, i.e. ${\vec{\cal V}}_0^+ = - i {\vec{U_0}}$ or ${\vec{\cal V}}_0^- = {\vec k} - i \vec\nabla \ln ( _1 F_1 (\vec k , \vec r) )$, respectively, with $_1F_1 (\vec k , \vec r) =~_1F_1(iZ/k,1,ikr-i\vec k.\vec r)$. Using Eqs. (\ref{eq:Newton2}) and (\ref{eq:NewtonStationnary}), the equation for the generalized action reads:
\begin{equation}
\frac{\partial {\cal S}}{\partial t} = - \vec{r}.\vec{E}(t) - \vec{A}(t).{\vec{\cal V}}_0^{\pm} - \frac{A(t)^{2}}{2}
\end{equation}
which temporal integration provides:
\begin{equation}
{\cal S}(\vec r, t) = {\cal S}_0^{\pm} + \vec{A}(t).\vec{r} - \vec{\alpha}(t).{\vec{\cal V}}_0^{\pm} - \int_{t_0}^t dt' \frac{A(t')^{2}}{2}
\label{eq:generalizedAction}
\end{equation}
where ${\cal S}_0^{\pm}$ is the action associated with the initial unpertubed state, and $\vec{\alpha}(t) = \int_{t_0}^t dt' \vec{A}(t')$. Since $\exp (i {\cal S}_0^{\pm}) = \Phi_0^{\pm} (\vec r, t)$, with $\Phi_0^{+} = \Phi_0$ and $\Phi_0^{-} = \Phi_k^{-}$, the wave function reads:
\begin{equation}
\chi^{\pm}(\vec{r},t) = \Phi_0^{\pm}(\vec r , t) \exp \left( i \vec{A}(t).\vec r - i \vec{\alpha}(t).{\vec{\cal V}}_0^{\pm} - \frac{i}{2} \int_{t_0}^t dt' A(t')^2 \right)
\label{eq:chiSRT}
\end{equation}
The introduction of the latter expression into the TDSE leads to a residual term which now depends on $\vec \alpha (t)$. This state is hereafter called $\chi_{i,f}^{\pm(2)SRT}(\vec{r},t)$.

The crossed term $\vec{\alpha}(t).{\vec{\cal V}}_0^{\pm}$, which depends on both time and space, now appears in the Volkov phase. This is a new theoretical expression for a laser-dressed wave function which emerges from the fractal structure of space, and is symmetric whatever the continuum or bound nature of the state. We emphasize that a velocity of the initial unperturbed bound state can be defined, allowing one to reintroduce the concept of MTA. Since ${\vec{\cal V}}_0$ is linked to the gradient of the unperturbed state, this term appears related to the operator $\exp (-\vec{\alpha}(t).\vec \nabla)$ which suggests an exponential solution of Eq. (\ref{eq:dfdtsplitted}). This is addressed in the following Section.

\section{Laser-dressed atomic states within standard quantum mechanics using SRT suggestions}
\label{sec:backToQM}
\subsection{Derivation for laser-dressed bound states}
The derivation of laser-dressed bound states cannot be performed in the same way as previously because the integration of Eq. (\ref{eq:dpdt}) has no meaning in QM for a bound electron: the value of the initial momentum $\vec{p}(-\infty)$ is not defined for a stationary bound state. However, starting from the same ansatz for the form of the approximate wave function as in Eq. (\ref{eq:chif}) where $f^-$ is substituted by $f^+$, and making the assumption that $\vec{p}$ is the expectation value of the momentum, it can be shown that $\vec{p}(t) = \vec{A}(t)$ anticipating that the $f^+$ function to be determined is the unperturbed initial bound wave function (see \ref{sec:appendix}). With this new assumption, still using $\partial \vec{p}(t)/ \partial t = - \vec{E}(t)$, Eq. (\ref{eq:TDSE}) leads to:
\begin{eqnarray} 
\label{eq:dfdtsplitted2}
i\frac{\partial f^+(\vec{r},t)}{\partial t} = - i \vec{A}(t).\vec{\nabla} f^+(\vec{r},t) \\
\left( -\frac{\nabla^{2}}{2} - \frac{Z}{r} \right) f^+ (\vec{r},t) = 0
\label{eq:dfdtsplitted2bis}
\end{eqnarray}
Eq. (\ref{eq:dfdtsplitted2bis}) is not physically relevant because, to correspond to a Schr\"odinger equation, at least an energy term accounting for the eigen value problem is missing. Within this procedure, the information on the energy of the initial unperturbed bound wave function has been lost due to the zero initial value of the momentum, whereas it has succeeded for the continuum state since $\vec{p}_0 = \vec k$, leading to the consistent energy $E_0 = p_0^2/2 = k^2/2$ appearing in the Volkov phase. We can get around this problem by making a new ansatz on the approximate wave function including the eigenenergy $E_i$ of the initial unperturbed state, as introduced in the SRT derivation based on (\ref{eq:ansatzChii}) where the kinetic energy is substituted by the total electron energy, i.e. $p^2/2 \rightarrow p^2/2 + E_i$. The new ansatz then reads:
\begin{equation} 
\chi_i^+(\vec{r},t) = f^+(\vec{r},t) \times \exp \left( i\vec{p}(t).\vec{r} - i \int_0^t dt' \left( \frac{p^{2}(t')}{2} + E_i \right) \right)
\label{eq:chii}
\end{equation}
With the same level of approximation as previously, in particular $\partial \vec p / \partial t = - \vec E (t)$ and $\vec p (t) = \vec A (t)$ ($\vec{p}_0 = \vec{0}$), the introduction of the expression (\ref{eq:chii}) in Eq. (\ref{eq:TDSE}) leads to:
\begin{eqnarray} 
\label{eq:dfdtsplitted3}
i\frac{\partial f^+(\vec{r},t)}{\partial t} = - i \vec{A}(t).\vec{\nabla} f^+(\vec{r},t) \\
\left( -\frac{\nabla^{2}}{2} - \frac{Z}{r} \right) f^+ (\vec{r},t) = E_i f^+ (\vec{r},t)
\end{eqnarray}
Still neglecting the right hand term $\vec{A}.\vec{\nabla} f^+$ of Eq. (\ref{eq:dfdtsplitted3}), $f^+$ does not depend on time, and therefore is nothing but the spatial part of the initial unperturbed bound state, $\varphi_i(\vec r)$. The bound CV state then reads:
\begin{equation} 
\chi_i^{+(1)} = \Phi_i(\vec{r},t) \times \exp \left( i\vec{A}(t).\vec{r} - \frac{i}{2} \int_{0}^t dt' A^{2}(t') \right)
\label{eq:solchi+}
\end{equation}
We emphasize that this derivation has been possible anticipating that $f^+ = \varphi_i$ to evaluate the expectation value of the momentum. Also note that the level of approximation between continuum and bound CV state is slightly different since the neglected terms can be written as $\vec{A}.\vec{\nabla} (f^- \exp(-i\vec k . \vec r))$ and $\vec{A}.\vec{\nabla} f^+$, respectively. An equivalence is obtained only for $\vec k = \vec 0$ which indeed corresponds to the zero expectation value of momentum of the bound state ($\vec{p}_0 = \vec 0$). Based on this kind of derivation, a consequence is that the continuum CV state includes the crossed term $\vec{k}.\vec{\alpha}(t)$ in its Volkov phase, whereas this is not the case for bound CV states.

\subsection{A more general derivation}
The following derivation has been suggested by the SRT framework due to the crossed term $\vec{\alpha}(t).{\vec{\cal V}}_0^{\pm}$, appearing in the Volkov phase of Eq. (\ref{eq:chiSRT}), which is related to the quantum operator $\vec{\alpha}(t).\vec \nabla$. The derivation is provided here for a continuum state. Let us restart from the system made of Eqs. (\ref{eq:dfdtsplitted}) and (\ref{eq:dfdtsplittedn}) to be solved. Instead of neglecting the crossed term $\vec{A}(t).\vec{\nabla} f^-(\vec{r},t)$ in Eq. (\ref{eq:dfdtsplitted}), the latter can be solved exactly with the method of characteristics leading to $f^-(\vec r, t) = f_0^-(\vec r - \vec\alpha (t))$ where $f_0^-(\vec r)$ is, for the moment, a general function only depending on space which corresponds to the initial condition. An expansion of $f_0^-(\vec r - \vec\alpha (t))$ around $\vec r$ leads to:
\begin{equation}
f_0^-(\vec r - \vec\alpha (t)) = \sum_{n=0}^\infty \frac{(-\vec\alpha (t))^n \vec \nabla^n f_0^-(\vec r)}{n!}
\label{eq:expansion}
\end{equation}
which also can be written as:
\begin{equation}
f^-(\vec r, t) = \exp (-\vec \alpha(t) . \vec\nabla) f_0^-(\vec r)
\end{equation}
because the expansion of the exponential function provides the n-order spatial derivative appearing in Eq. (\ref{eq:expansion}), bridging with the previous SRT considerations about the $\vec{\alpha}(t).{\vec{\cal V}}_0^{\pm}$ phase. As previously, $f_0^-(\vec r)$ can be determined as a solution of Eq. (\ref{eq:dfdtsplittedn}), i.e. $f_0^-(\vec r) \propto _1F_1(iZ/k,1,ikr-i\vec k.\vec r)$. Note that $f^-(\vec r, t) = f_0^-(\vec r - \vec\alpha (t))$ is not an exact solution of Eq. (\ref{eq:dfdtsplittedn}). Indeed, making the transformation $\vec r \rightarrow \vec r + \vec\alpha (t)$, it transforms into:
\begin{equation}
\left( -\frac{\nabla^{2}}{2} - \frac{Z}{\| \vec r + \vec\alpha (t)\| } -i \vec{k}.\vec{\nabla} \right) f_0^- (\vec{r}) = ( \vec\alpha (t).\vec\nabla V_c(\vec r) + \frac{\vec\alpha (t)^2}{2!}\vec\nabla^2 V_c(\vec r) + ... ) f_0^- (\vec{r})
\label{eq:hypergeo}
\end{equation}
which the right hand term is different from zero in general. Finally, within this procedure, an approximate solution of the TDSE reads:
\begin{eqnarray}
\chi_f^{-(2)}(\vec{r},t) & = \exp \left( i\vec p (t).\vec r - \frac{i}{2} \int_{\infty}^{t} dt \vec p (t)^2  -\vec \alpha (t) . \vec\nabla \right) f_0^-(\vec r) \\
& = \Phi_{\vec{k}}^- (\vec{r}-\vec\alpha,t) \times \exp \left( i\vec A (t).\vec r - \frac{i}{2} \int_{\infty}^{t} dt \vec A (t)^2  \right) \nonumber
\end{eqnarray}
with $\vec p (t) = \vec k + \vec{A}(t)$. The transformation $\vec r \rightarrow \vec r + \vec\alpha (t)$ in the Coulomb state, equivalent of applying the phase $-\vec\alpha (t).\vec\nabla$ to the wave function, is nothing but considering the electron in the moving Kramers-Hennenberger (KH) reference frame \cite{Henneberger, Kramers}. This frame is used to write the TDSE in the so-called acceleration gauge. This form of wave function was also obtained in \cite{Faisal, Macri2003} with different derivations. 

By analogy, following previous derivations, a bound laser-dressed state can be constructed:
\begin{eqnarray} 
\label{eq:solchi+2}
\chi_i^{+(2)}(\vec{r},t) & = \exp \left( i\vec{A}(t).\vec{r} - \frac{i}{2} \int_{0}^t dt' A^{2}(t') -\vec \alpha (t) . \vec\nabla \right) \Phi_i(\vec{r},t) \\
& = \Phi_i(\vec{r}-\vec\alpha (t),t)\exp \left( i\vec{A}(t).\vec{r} - \frac{i}{2} \int_{0}^t dt' A^{2}(t') \right) \nonumber
\end{eqnarray}
Doing so, this expression is the exact analogue of $\chi_f^{-(2)}$ where the continuum unperturbed state is substituted by the bound one. Note however that as in the previous section, the derivation is not analogous because some further assumptions are required for the bound state.

\section{Discussion and application to ionization}
\label{sec:discussion}
The next section \ref{subsec:comparison} is devoted to the analysis and comparison of the various obtained laser dressed wave functions. For the states derived within the QM framework, the beginning of the discussion stresses on the link between the presence of the quiver motion $\alpha$ (pertaining to the acceleration gauge) and the sudden approximation. Details are intentionally examined because the discussion should also stand for the SRT-derived states including the quiver motion since the mathematical structure of both QM and SRT laser dressed wave functions is similar. These considerations allow us to define physical conditions for a first simple application of the presently derived states. That also permits to get more insight on the reliability of these wave functions. Section \ref{subsec:application} thus addresses the ionization of hydrogen targets by attosecond XUV pulses within the sudden approximation where SRT and QM predictions are compared.

\subsection{Comparison of the various wave functions}
\label{subsec:comparison}
In order to clearly highlight the difference in between the wave functions, they are all summarized hereafter:
\begin{eqnarray}
\chi_f^{-(1)}(\vec{r},t) & = \Phi_{\vec{k}}^- (\vec{r},t) \times \exp \left( i\vec{A}(t).\vec{r} - i\vec{k}.\vec{\alpha}(t)  - \frac{i}{2} \int_\infty^t dt' A^{2}(t') \right) \\
\chi_f^{-(2)}(\vec{r},t) & = \Phi_{\vec{k}}^- (\vec{r}-\vec\alpha,t) \times \exp \left( i\vec A (t).\vec r - \frac{i}{2} \int_\infty^t dt' \vec A (t')^2  \right) \\
\chi_i^{+(1)}(\vec{r},t) & = \Phi_i(\vec{r},t) \times \exp \left( i\vec{A}(t).\vec{r} - \frac{i}{2} \int_{0}^t dt' A^{2}(t') \right) \\
\chi_i^{+(2)}(\vec{r},t) & = \Phi_i(\vec{r}-\vec\alpha,t) \times \exp \left( i\vec{A}(t).\vec{r} - \frac{i}{2} \int_{0}^t dt' A^{2}(t') \right) \\
\chi_i^{+ (1)SRT}(\vec{r},t) & = \Phi_0(\vec r , t) \exp \left( i \vec{A}(t).\vec r - \frac{i}{2} \int_0^t dt A(t)^2 \right) \\
\chi_{i,f}^{\pm (2)SRT}(\vec{r},t) & = \Phi_0^{\pm}(\vec r , t) \exp \left( i \vec{A}(t).\vec r - i \vec{\alpha}(t).{\vec{\cal V}}_0^{\pm} - \frac{i}{2} \int_{t_0}^t dt' A(t')^2 \right)
\end{eqnarray}
We start the analysis on the bound CV states obtained with the standard QM, i.e. $\chi_i^{+(1)}$ and $\chi_i^{+(2)}$. $\chi_i^{+(1)}$ is used as a reference since it is the simplest expression. $\chi_i^{+(2)}$ clearly appears as a generalization of $\chi_i^{+(1)}$ since $\lim_{\alpha \to 0} \chi_i^{+(2)} = \chi_i^{+(1)}$. Indeed it accounts for more physical processes a priori since it also includes the change in the reference frame of the Coulomb state due to the laser electric field; i.e. the wave function includes a phase which pertains to the KH transformation \cite{Henneberger, Kramers}. It then turns out that this wave function form accounts for the electron dynamics \textit{simultaneously} in both the coulomb and the laser electric fields. $\chi_i^{+(1)}$, which simply consists of the product of a Coulomb state and the Volkov phase, cannot account for this effect \cite{Smirnova2006, Smirnova2008}. Effectively it has been shown that the introduction of the quiver motion provides better results \cite{Gravielle2012}. Since it accounts for the KH frame, i.e. the acceleration gauge, this CV state keeps in memory the felt acceleration during the laser interaction, thus leading to a wave function different from the initial one even if $\vec A$ vanishes by the end of the interaction at $t=\tau$, i.e. $\chi_i^{+(2)} (\vec{r},t=\tau) = \Phi_i (\vec{r} - \vec \alpha (t=\tau),t=\tau)$ whereas $\chi_i^{+(1)} (\vec{r},t=\tau) = \Phi_i (\vec{r},t=\tau)$ (the same reasoning applies for ingoing waves). Therefore, a memory of the interaction history remains when $\alpha (t=\tau) \neq 0$.

When $\chi_i^{+(2)}$ is introduced in the TDSE, the residual term at the lowest order reads $\vec \alpha . \vec \nabla V_c$. It corresponds to the energy associated with the action of the coulomb force over the electron displacement $\vec \alpha$ (see \ref{sec:appendix2} for its classical derivation). This expression appears contradictory since the coulomb field should also influence on the electron displacement. It turns out that this expression makes sense if the coulomb force does not influence the electron trajectory at all, i.e. the displacement is only induced by the external laser electric field. This case corresponds to a very intense laser pulse where the dynamical influence of the nucleus can be neglected, or very short interaction time, shorter than the orbital period, $T_n$, for which the electron has not enough time to adapt to the external perturbation, i.e. the coulomb wave function remains the same during the whole interaction. The latter condition corresponds to the sudden approximation which is all the better that the pulse duration is short, in particular $\tau \ll T_n$ is required \cite{Reed1990, Forre2005}. This analysis suggests that under such conditions, to evaluate a transition amplitude, an expression including temporal integrations as in the perturbative approach is no longer required to account for the temporal dynamics of the interaction. The direct projection of $\chi_i^{+(2)}$ onto a final unperturbed state may provide a reliable expression of the transition amplitude. It is done and analysed in Section \ref{subsec:application}.

To go further in the analysis of conditions where these wave functions should provide reliable predictions, the introduction of $\chi_i^{+(2)}$ in the TDSE actually leads to an infinite serie of residual terms as exhibited by Eq. (\ref{eq:hypergeo}), which may be written as an exponential function. Compared to $\chi_i^{+(1)}$ which only provides one residual term in the TDSE as $-\vec A . \vec \nabla \Phi_0$, the domain of validity of $\chi_i^{+(2)}$ is not so clear (possibly, a divergence may appear due to the infinite sum). As a compromise, the wave function can be truncated to only provide a first order coulomb correction in the laser-induced electron dynamics and leads to a finite number of residual terms in the TDSE which the amplitude may be more easily managed (see \ref{sec:appendix3}). Considering an s state, since $|\nabla \Phi_i|\propto 1/n$ where $n$ is the principal quantum number, the more excited the target, the better the CV approach \cite{D2}. More detailed conditions of reliability of these states are provided in \cite{Faisal}.

Considerations about the quantum-mechanical continuum states are equivalent to the previous one. However, the phase term $\vec{k}.\vec{\alpha}(t)$ of $\chi_f^{-(1)}$ corresponds to the frame translation only applied to the plane wave part of the Coulomb wave function. This phase then appears as a spurious term since one may have $\lim_{\alpha \to 0} \chi_f^{-(2)} = \Phi_{\vec{k}}^- (\vec{r},t) \times \exp \left( i\vec{A}(t).\vec{r} - \frac{i}{2} \int_\infty^t dt' A^{2}(t') \right)$ where the phase $\vec{k}.\vec{\alpha}(t)$ does not appear. It is only correct in the case $k=0$ which actually pertains to the bound case for which the quantum-mechanical velocity is zero.

\vspace{0.5cm}

We now consider the CV states derived from the SRT framework.
In addition to the standard Volkov phase of the $\chi_f^{\pm(1)}$ reference state, $\chi_{i,f}^{\pm (2)SRT}(\vec{r},t)$ includes a term $\exp (-\vec \alpha (t) . \vec \nabla \ln \Phi_0^{\pm}(\vec r , t) )$. An analogy with the term $\exp (-\vec \alpha (t) . \vec \nabla ) \Phi_0 (\vec r , t)$ of $\chi_i^{+(2)}$ and $\chi_i^{+(2)SRT}$ clearly appears, showing that $\chi_i^{+(2)SRT}$ also pertains to the acceleration gauge. After expansion of the exponential function, the main difference lies in the fact that $\chi_{i,f}^{\pm (2)SRT}(\vec{r},t)$ consists of powers of $\vec \nabla  \Phi_0$ (or ${\vec{\cal V}}_0$), whereas $\chi_i^{+(2)}$ includes the high-order derivatives of $\Phi_0$. Up to the first order, both description are strictly equivalent. Considering all orders, in case of 1s state targets where $\Phi_0$ is a pure exponential function, both descriptions appears equivalent. A departure from this conclusion is expected for excited states. For ns states, since ${\cal V}_0 \propto 1/n$, the expansion of $\chi^{+ (2)SRT}(\vec{r},t)$ in powers of ${\cal V}_0$ should converge rapidly. Note also that the wave function normalization is changed within the SRT framework since the phase $-i \vec{\alpha}(t).{\vec{\cal V}}_0^{\pm}$ includes a real part depending on space (this is further discussed in the following section).

The fractal framework also allows us to clarify the physical meaning of the residual term of the TDSE by now writing the neglected crossed term as ${\vec{\cal V}}_0.\vec{A}$: it corresponds to the projection of the initial velocity field onto the direction of laser polarization. In the case of an initial s state (exhibiting a spatial spherical symetry, i.e. with ${\vec{\cal V}}_0$ only including a radial component) and a linearly polarized laser pulse, the crossed term accounts for the spatial symetry breaking of the initial state during the interaction. Considering an initial state with a large angular momentum, ${\vec{\cal V}}_0$ is mainly oriented in a given direction. The irradiation of such prepared targets by a pulse polarized in the direction perpendicular to the main of ${\vec{\cal V}}_0$ then would decrease the amplitude of the crossed term, and thus increases the accuracy of the CV bound state.

\subsection{Application to ionization by XUV pulses}
\label{subsec:application}
This section is devoted to a first application to evaluate whether the wave functions including the quiver motion $\vec \alpha$ are able to make physically reasonable predictions and thus get a further insight on their reliability. The previous considerations have shed light on the possible applications of these CV wave functions due to the shift in the reference frame. Within the sudden approximation, the fact that this kind of states may be used to predict transition amplitude by directly projecting it onto a final unperturbed state has been discussed above. Such physical conditions correspond to the irradiation of atoms by attosecond XUV pulses which is thus chosen here as a first application due to its simplicity. Note that studies within these conditions have been carried out in \cite{Dimitrovski2005, Yudin2007, Smirnova2006} for instance.

These physical conditions roughly corresponds to $\omega > 1 a.u.$ and $\tau < 10 a.u.$ for H(1s) targets, whatever the laser intensity. A longer pulse duration could be used for excited atomic targets having a longer orbital period, then still satisfying the sudden approximation. The differential ionization probability is given by:
\begin{equation}
\frac{dP}{d\vec k} = \left| < \Phi_{\vec{k}}^- \ | \ \chi_i^{+}(t=\tau) > \right|^2
\end{equation}
where the bound outgoing laser-dressed state is used. Similar calculations could be performed by using the ingoing wave functions. If $\chi_i^{+} = \chi_i^{+(1)}$, then $P=0$ since initial and final states are orthogonal ($A(\tau) = 0$). Note that even if $\alpha \neq 0$, the phase $\vec k.\vec \alpha$ cannot induce any transition since it does not depend on space. For the present purpose devoted to exhibit the ability of the presently obtained CV states to capture the ionization process within these extreme conditions, determining an exact expression of $\alpha$ is not necessary. For given conditions, it is equal to $E_0 \tau / \omega$ for instance.

\vspace{0.5cm}

Using the SRT framework to derive the laser-dressed wave function, the differential probability is given by:
\begin{equation}
\frac{dP}{d\vec k} = \left| \int d\vec r \Phi_{\vec{k}}^{-} (\vec r)^* e^{- i {\vec{\cal V}}_0^{\pm}.\vec{\alpha}(\tau)} \Phi_0(\vec r)  \right|^2
\label{eq:P_SRT}
\end{equation}
where non contributing phases have been removed. A direct analytical evaluation of this spatial integral appears complicated in the general case, i.e. whatever the target and the value of $\alpha$. To simplify the analytical calculations and highlight the main physical processes, we choose the limit cases $\alpha \ll 1$ and $\alpha \gg 1$ for the present analysis. For the former, the exponential function of Eq. (\ref{eq:P_SRT}) can be expanded in powers of $- i {\vec{\cal V}}_0^{\pm}.\vec{\alpha}(\tau)$ with ${\vec{\cal V}}_0 = -i \vec \nabla \ln \Phi_0$. The zero-order also does not contribute since $\Phi_{\vec{k}}^-$ and $\Phi_0$ are orthogonal. At the first order, one gets:
\begin{equation}
\frac{dP}{d\vec k} = \left| \int d\vec r \Phi_{\vec{k}}^- (\vec r)^* {\vec{\cal V}}_0^{\pm}.\vec{\alpha}(\tau) \Phi_0(\vec r)  \right|^2 = \left| \int d\vec r \Phi_{\vec{k}}^- (\vec r)^* \vec{\alpha}(\tau).\vec \nabla \Phi_0(\vec r) \right|^2
\label{eq:P_SRT_O1}
\end{equation}
This approach is able to capture the one-photon absorption process since the scalar product $\vec \alpha(\tau).\vec\nabla$ provides a spherical harmonics $Y_l^m$ with $l=1$ and $m=0$ leading to the angular momentum selection rules through the angular integration. The probability as predicted by the Born first-order standard perturbation theory in the velocity gauge reads:
\begin{eqnarray}
\label{eq:P_Born}
\frac{dP}{d\vec k} & = \left| \ -i \int_0^\tau dt \int d\vec r \ \Phi_{\vec{k}}^- (\vec r ,t)^* \ \vec A(t).\vec\nabla \Phi_0(\vec r,t) \ \right|^2 \\
& = \left| \ \int_0^\tau dt A(t) e^{-i(E_0 - E_k)t} \int d\vec r \ \Phi_{\vec{k}}^- (\vec r)^* \ \hat{e}.\vec\nabla \Phi_0(\vec r) \ \right|^2 \nonumber
\end{eqnarray}
where $\hat{e}$ is the unitary vector oriented in the laser polarization direction. Writing $A(t)=A_0 e^{-i\omega t}$, if $\omega \gg E_k - E_0$, the ionization probability (\ref{eq:P_SRT_O1}) is similar to the expression (\ref{eq:P_Born}) since $\alpha = \int dt A(t)$. This condition corresponds to the non-resonant one-photon absorption: only the part of the energy spectrum of the ejected electrons which is sufficiently far from the energy conservation $E_k = E_0 + \omega$ should be considered. The ionization probability in this region of the spectrum is non-zero due to the finite duration of the laser pulse. A similar conclusion has been obtained by evaluating a matrix element in the Kramers-Henneberger frame \cite{Henneberger}. At second-order, based on the same considerations as previously with an expansion to the second order, the non-resonant two-photon absorption is included in the expression (\ref{eq:P_SRT}). That can also be generalized to the n-photon absorption process which requires non negligible values of $\alpha$ allowing one significant transition probability even for high order transitions.

The case $\alpha \gg 1$ is now considered. $\alpha$ increasing (or the electric field), a transition from the multiphoton absorption to the tunneling regime is expected, ultimately leading to a full ionization, i.e. $P=1$. Let us consider now the full $\chi_{i,f}^{\pm (2)SRT}$ state without any particular development. The argument of the dressing exponential function is $-i \vec \alpha . \vec {\cal V}_0$. To simplify the present discussion but without any loss of generality, the simplest H(1s) state is considered, i.e. $\vec {\cal V}_0 = i\hat{r}$, leading to the exponential argument $\alpha \cos \theta$ in spherical coordinates where the laser polarization is oriented along the $z$ axis. Since this argument is real, this dressing of the initial state changes the wave function normalization. Within extreme conditions where $\alpha$ becomes significantly large, the norm may even diverge leading to an ionization probability larger than unity. This behavior can be understood by writing the laser dressed velocity field. From $\vec {\cal V} = \vec \nabla {\cal S}$ and using Eq. (\ref{eq:generalizedAction}), we get $\vec {\cal V} = \vec {\cal V}_0 + \vec A (t) - \vec \nabla (\vec\alpha . \vec {\cal V}_0)$. When $\alpha$ increases, there is no limit to the increase in the velocity amplitude, leading to the above-mentioned possible divergence. It is due to the fact that the SRT exhibits classical features in the sense that its fundamental equation is a Newton-like one. The velocity thus may reach any arbitrary large value, larger that the speed of light in particular. We suggest that a limitation of the generalized velocity amplitude thus could be introduced by using a relativitic version of the presently presented SRT. It can be achieved by further assuming that, in addition to space, the time also exhibit a fractal structure \cite{nottale1993towards}. Doing so, any possible divergence is removed, making the present approach more reliable through this renormalization procedure. This development is out of the scope of the present paper and will be adressed elsewhere. It is worth noting that this change in the normalization of the wave function does not have any influence in the case $\alpha \ll 1$. For moderate values of $\alpha$ in between both asymptotic regimes, the SRT-derived states still should provide reliable ionization probabilities but different from the one obtained from the QM, it is addressed hereafter.

\vspace{0.5cm}

By using the expression $\chi_i^{+(2)}$ obtained within QM, the differential probability reads:
\begin{equation}
\frac{dP}{d\vec k} = \left| \ \int d\vec r \ \Phi_{\vec{k}}^- (\vec r)^* \ \Phi_0(\vec r - \vec \alpha(\tau)) \ \right|^2
\label{eq:P_QM}
\end{equation}
Note that a similar expression has be obtained in \cite{Klaiber2015}. However the nature of the wave function spatial shift is different since it is interpreted as the light pressure within the framework of nondipole effects.

A direct analytical evaluation of the spatial integral of Eq. (\ref{eq:P_QM}) also appears complicated in the general case, i.e. whatever the value of $\alpha$. To simplify the analytical calculations, highlight the main physical processes, and compare to the SRT predictions, we choose again the limit cases $\alpha \ll 1$ and $\alpha \gg 1$. For $\alpha \ll 1$, $\Phi_0(\vec r - \vec \alpha(\tau))$ can be expanded as shown in Eq. (\ref{eq:expansion}). The zero-order still does not contribute. At first order, one gets:
\begin{equation}
\frac{dP}{d\vec k} = \left| \ \int d\vec r \ \Phi_{\vec{k}}^- (\vec r)^* \ \vec \alpha(\tau).\vec\nabla \Phi_0(\vec r) \ \right|^2
\label{eq:P_QM2}
\end{equation}
which is exactly the same expression as (\ref{eq:P_SRT_O1}). This laser dressed wave function thus also allows one to account for the non-resonant one-photon absorption process. That can also be generalized to the non-resonant n-photon absorption process which requires non negligible values of $\alpha$ allowing one significant transition probability even for high order transitions. However some discrepancies between SRT and QM predictions should appear because pre-factors of each power of the field amplitude obtained from the exponential expansion are different, which is linked in particular to the different normalization of states. Asymptotically for $\alpha \gg 1$, we may assume that $\Phi_{\vec{k}}^- = \exp (i\vec k . \vec r)/(2\pi)^{3/2}$ since the integrand contributes mainly around $\vec r=\vec \alpha$, i.e. $kr\rightarrow \infty$. In that case, note that the ionization probability is nothing but the Fourier transform of the initial state. The expression (\ref{eq:P_QM}) can be evaluated analytically for H(1s) targets for instance, one gets:
\begin{equation}
\frac{dP}{d\vec k} = \frac{8}{\pi^2 (1+k^2)^4}
\label{eq:dPdk}
\end{equation}
An isotropic distribution is obtained. Considering the present physical system, that may be surprising at first glance because the laser linear polarization would break the symmetry of the initial state during the ionization process. That does not take place due to the very short interaction time. In this situation, the whole initial electron cloud is shifted in space on a timescale which does not allow any influence of the coulomb field on the electron dynamics. This situation corresponds to the barrier-suppression regime \cite{Bauer}. In case of pulse duration longer than the orbital period, the transition becomes adiabatic, the electron then may experience the coulomb field during the ionization process, leading to the breaking of the initial energy distribution symetry.

We have checked that integrating the differential probability (\ref{eq:dPdk}) over the momentum leads to unity. To go further, whatever the value of $\alpha$, a numerical evaluation of the expression (\ref{eq:P_QM}) using the Maple software has been performed (for H(1s) with $k=1$). The previously discussed asymptotic behaviors are retrieved. By plotting this ionization probability as a function of $\alpha$, the curve can be very correctly fitted with the normalized expression $\exp ( - 3 / 10 \alpha )$ with an error $\chi^2 \simeq 4 \times 10^{-3}$. The latter expression accounts for tunneling ionization and barrier-suppression regime asymptotically \cite{Bauer}. Thus that confirms the previous considerations on the ability of the present approach to capture this process. Note that the present ionization probability depends implicitely on the laser frequency $\omega$ through the electron quiver motion $\alpha (\tau)$. The mathematical expression for the dependence on $\omega$ may appear different from previous theoretical developments due to different interaction regimes. It is noteworthy that the use of a transition amplitude form including a temporal integration, as $-i \int dt < \Phi_{\vec{k}}^- (t) \ | \ V_L (t) \ | \ \chi_i^{+}(t) > $, allows one to account for resonant multiphoton absorption \cite{Faisal}. Finally, since the SRT-derived state exihibits a similar mathematical structure, similar predictions are expected when the possible divergent behavior will be fixed through the above-mentioned procedure.

\section{Conclusion}
\label{sec:conclusion}
The present work has been motivated by the fact that using the momentum-translation approximation on initial bound states to construct laser-dressed bound wave functions is not well appropriate. The main goal of this paper was then to establish a new theoretical framework to derive laser-dressed wave functions. It is based on a fractal geometry of space renders it possible to properly define a momentum for a bound state. A bound Coulomb-Volkov has then be derived naturally as for continuum states. By using the most general ansatz for the wave function form, based on the eikonal including the action, a general derivation of CV states has been proposed, whatever the ingoing or outgoing nature of the initial state. A new expression for the laser-dressed state has been obtained, which was not derived from quantum mechanical derivations. This expression, due to its relation to the acceleration gauge, includes the electron quiver motion as a crossed term in the Volkov phase which then contains more information on the interaction history. Regarding in particular the bound state, this additional term includes the velocity field of the unperturbed bound wave function. That is a direct consequence of providing space with a fractal structure, i.e. it is a signature of this theoretical framework. This approach also more naturally sheds light on a neglected term during the derivation, then predicting that the CV wave function should be more accurate for well prepared initial states with a large angular momentum.

The previous developments have paved the way to revisit the derivation of approximate laser-dressed atomic wave functions within the framework of the standard theory of quantum mechanics. In case of an initial bound atomic state, a new ansatz has been proposed for the form of the laser-dressed wave function, allowing to derive formally a bound CV state following the same procedure as for the continuum, however with an additional assumption in the demonstration. A more general CV state including an additional phase pertaining to the Kramers-Hennenberger transformation has been derived still within the framework of the momentum-translation approximation.

Regarding both SRT and QM states which pertain to the acceleration gauge including the quiver motion, they are particularly well adapted to predict electron transitions even on a very short timescale where the dynamical influence of the coulomb potential is negligible. This situation corresponds to the sudden approximation where a transition amplitude can be evaluated by directly projecting a laser-dressed wave function onto a final ionized state. As a first application and test of the validity of these wave functions, within these conditions, calculations have been performed for the ionization of hydrogen targets by attosecond XUV pulses. Despite basic CV state predict no transition within these conditions, the present results based on the obtained generalized laser-dressed states are in a good agreement with previous results based on other approaches including an integration over time (permitting to follow the electron dynamics). In particular, the present simple approach allows to make predictions in various regimes depending on the laser intensity, going from the non-resonant multiphoton absorption to tunneling and barrier-suppression ionization for the largest intensities.

However in terms of application as the first proposed here, the SRT-derived states have not really provided more input than the QM-derived one. Following the present introduction of this new theoretical framework in the field of the laser-atom interaction, a work to consider other applications is under progress. We are performing further analytical and numerical investigations to determine conditions where the SRT-derived state could improve the description of certain mechanisms and the accuracy of associated predictions for transition probabilities. Especially, they will be used in time-dependent version of the transition amplitude to account for the resonant multiphoton absorption. Also, more generally as shown in this paper, we believe that through the mathematical description it offers, the fractal-space framework may be used to simplify future studies and possibly open new ways within the well-established quantum mechanics theory by revisiting some developments or by making analogies. In particular for processes involving the tunneling mechanism \cite{SFA1, Frolov} where the quantum mechanical developments exhibit eikonal-like expressions with integration techniques relying on the saddle point from which imaginary times may emerge.

\vspace{1.cm}

Fabrice Catoire and Vladimir Tikhonchuk are gratefully acknowledged for their comments on the present manuscript which have helped me to improve it. Bruno Dubroca is also gratefully acknowledged for his lectures on fractals.

\appendix
\section{Expectation value of the momentum for bound states}
\label{sec:appendix}
Here the expectation value of the momentum is calculated using bound CV states under the framework of the standard theory of quantum mechanics. It reads:
\begin{equation}
<\vec p(t)> = -i \int d\vec r \chi_i^+ (\vec r,t)^* \vec \nabla \chi_i^+ (\vec r,t)
\label{eq:evp}
\end{equation}
where $\chi_i^+ (\vec r,t) = \Phi_i (\vec r,t)\exp ( i\vec{A}(t).\vec{r} - i \int_0^t dt' A^{2}(t')/2 )$. It follows that $\vec \nabla \chi_i^+ = \exp ( i\vec{A}(t).\vec{r} - i \int_0^t dt' A^{2}(t')/2 ) \nabla \Phi_i + i \vec A \Phi_i$. Eq. (\ref{eq:evp}) then transforms into:
\begin{equation}
<\vec p(t)> = -i \int d\vec r \Phi_i^* \vec \nabla \Phi_i + \int d\vec r \Phi_i^* \vec A \Phi_i
\end{equation}
Since $\Phi_i$ is normalized to unity, $\int d\vec r \Phi_i^* \vec A \Phi_i = \vec A$. Due to angular symetries, it can be shown that $\int d\vec r \Phi_i^* \vec \nabla \Phi_i = \vec 0$. The expectation value of the momentum is then:
\begin{equation}
<\vec p(t)> = \vec A (t)
\end{equation}

\section{Evaluation of the classical action of the coulomb force during the laser-induced electron displacement}
\label{sec:appendix2}
Here is evaluated the energy associated with the action of the coulomb force over the laser-induced electron displacement. For this purpose, the classical equation for the electron dynamics is generalized by introducing the coulomb force:
\begin{equation}
\frac{\partial \vec{v}(t)}{\partial t} = - \vec{E}(t) - \vec \nabla V_c
\label{eq:eqAppendix2_1}
\end{equation}
where $m=1$. Considering a bound state with $\vec v_0 = 0$, multiplying by $\vec{v}(t)$ and assuming that $\vec{v}(t) \simeq \vec{A}(t)$ in the right hand side of Eq. (\ref{eq:eqAppendix2_1}), an integration over time provides the kinetic electron energy:
\begin{equation}
\frac{v(t)^2}{2} = \frac{A(t)^2}{2} - \vec \alpha (t) . \vec \nabla V_c
\label{eq:eqAppendix2_2}
\end{equation}
since $\int dt E(t)A(t) = A(t)^2 / 2$. The latter term is associated with the classical ponderomotive energy. The term $\vec \alpha (t) . \vec \nabla V_c$ accounts for the action of the coulomb force.

\section{A truncated wave function}
\label{sec:appendix3}
The fact that the residual term in Eq. (\ref{eq:hypergeo}) consists of an infinite sum of terms leaves unclear the level of accuracy of this wave function, even if small values of $\alpha (t)$ should provide a rapidly converging expansion in the sense of a perturbative expansion. We then propose an alternative scheme by only considering the first order of the expansion relative to the KH operator, i.e. $\exp (-\vec\alpha (t).\vec\nabla) \simeq 1 -\vec\alpha (t).\vec\nabla$. Another expression of a laser-dressed state then reads:
\begin{equation}
\chi_f^{-(3)} = \exp \left( i\vec A (t).\vec r - \frac{i}{2} \int dt \vec A (t)^2 \right) (1 -\vec\alpha (t).\vec\nabla) \Phi_{\vec{k}}^- (\vec{r},t)
\end{equation}
$\chi_f^{-(1)}$ takes \textit{independently} into consideration the influence of the laser and coulomb electric fields because it is a simple product of the Volkov phase and the unperturbed coulomb state \cite{Smirnova2008}. $\chi_f^{-(3)}$ includes a first order correction to the standard CV state which now accounts for the coupled laser-coulomb influence on the electron dynamics.

The truncation of the expansion of $\exp (-\vec \alpha (t) . \vec\nabla)$ appearing in (\ref{eq:solchi+2}) to the first order provides the analogue of $\chi_f^{-(3)}$ for a bound laser-dressed state.

\end{document}